\documentstyle[preprint,eqsecnum,tighten,prb,aps]{revtex}  
\begin{document}
\draft \preprint{UBCTP-93-18} \title{ MONOPOLE-CATALYSED BARYON DECAY: A
BOUNDARY CONFORMAL FIELD THEORY APPROACH} \author{ Ian Affleck$^{a,b}$ and
Jacob Sagi$^b$} \address{$^{(a)}$Canadian Institute for Advanced Research and
$^{(b)}$Physics Department, University of British Columbia, Vancouver, B.C.,
V6T1Z1, Canada}  \date{\today} \maketitle \begin{abstract} Monopole-mediated
baryon number violation, the Callan-Rubakov effect, is reexamined using
boundary conformal field theory techniques.  It is shown that the low-energy
behaviour is described simply by free fermions with a conformally invariant
boundary condition at the dyon location. When the number of fermion flavours
is greater than two, this boundary condition is of a non-trivial type which
has not been elucidated previously. \end{abstract} \pacs{ }

\narrowtext \section{Introduction} \label{sec:intro}

A spectacular effect of the Adler-Bell-Jackiw anomaly is the catalysis of
baryon number violation at strong-interaction rates by dyons, electrically
charged magnetic monopoles.\cite{Rub,Call,Call-Das,Pol}  Since the important
physics occurs in the s-wave channel during scattering of fermions {}from the
dyon, the problem can be mapped into a (1+1)-dimensional one.  It was
observed that, when the number of massless flavours, $N$ obeys $N \leq 2$,
all the complicated physics of the fermion-dyon interaction can be reduced,
at low energies, to a simple boundary condition on otherwise free
fermions.\cite{Call-Das,Pol}  On the other hand, in the physically
interesting case $N>2$, in which fractional quantum-number production
occurs, it was found that the appropriate boundary conditions on the
currents did {\it not} correspond to any linear boundary condition on the
fermions.  Since this time, great progress has been made in studying
conformally-invariant boundary conditions in (1+1)-dimensional field
theories, due to seminal work by Cardy\cite{Cardy}.  Cardy's methods,
originally developed to study two-dimensional classical statistical systems
with boundaries, have been generalized to the study of quantum-impurity
problems.\cite{A,AL1,AL2,LA1,LA2}  It has become evident that, in a large
class of such problems, the low energy physics can be described without
explicit reference to the quantum impurity (in this case the dyon) by simply
imposing a conformally-invariant boundary condition.  Cardy made precise the
meaning of ``conformally invariant boundary condition'', through the device
of modular transformation and the concept of a ``boundary state''.  In
general these ``boundary conditions''cannot be written as local, linear
boundary conditions on the original fields in the problem.  In the case of
the Kondo problem, the linear boundary conditions correspond to  Fermi
liquid fixed points, while the other ones do not.

The purpose of the present work is to re-examine the Callan-Rubakov
baryon-number violation effect {}from a slightly more modern perspective.  We
will show explicitly that, for all values of $N$, the low-energy physics is
completely determined by conformally-invariant boundary conditions, \`a la
Cardy.  An essential tool for this construction is non-abelian bosonization;
the boundary conditions on the currents, derived
earlier,\cite{Call,Call-Das,Pol}  must be supplemented by appropriate
``gluing conditions'' governing the way that charge and flavour quantum
numbers are combined.  The topological vacuum angle appears as an explicit
free parameter in the boundary state.  This leads directly to a calculation
of the Green's functions in the vicinity of the dyon.  We hope that our
method may serve to clarify the physics of the Callan-Rubakov effect as well
as perhaps suggesting analogous phenomena in other areas of physics.  The
dyon problem requires an  extension of Cardy's boundary conformal field
theory techniques to deal with an infinite number of conformal towers.  It
turns out that the boundary state corresponding to the dyon is closely
related to one that occurs in the problem of transmission through barriers
in quantum wires.\cite{AW}

In the next section we briefly review some results\cite{Call,Call-Das,Pol} on
dyons.  In Section III we briefly review Cardy's boundary formalism and
deduce the boundary states and finite-size spectrum for the dyon problem for
arbitrary $N$.  In Section IV we calculate  Green's functions  in the
presence of the dyon, directly {}from the boundary states.  These agree
exactly with the results of Polchinksi obtained by another method.

\section{Dyon-catalysed baryon number violation}

The electrical degree of freedom of the dyon corresponds essentially to a
rigid rotator collective co-ordinate associated with gauge transformations.
A ``stripped down'' version of the model, proposed by Polchinski,\cite{Pol}
involves otherwise free, right-moving fermions interacting near the origin
with a rigid rotator.  The Hamiltonian is: \begin{equation} H =
-i\int_{-\infty}^{\infty}dx \sum_{k=1}^N\psi^{\dagger k}\left({d \over
dx}-i\alpha f(x)\right) \psi_k(x) + {1\over 2I}\Pi^2 \label{Ham}
\end{equation} Here $\alpha$ is the rigid rotator co-ordinate and $\Pi$ is
the conjugate momentum, \begin{equation} [\Pi ,\alpha ] = -i. \end{equation}
$I$ is the moment of inertia of the rotator and $f(x)$ is an even function
which vanishes for $|x|>r_0$ where $r_0$ is of order the dyon core size. It
can be assumed to have integral one since we are free to rescale $\alpha$
and $\Pi$ and hence $I$. $\alpha$ represents a collective co-ordinate of the
classical dyon solution and $\Pi$ corresponds to the electric field or
charge of the dyon. Here we work with right-moving fermions on the whole
real line.  This is obtained {}from the s-wave problem, originally defined on
the positive half-line, by reflecting the incoming wave (left-movers) to the
negative axis.  A crucial and subtle point in the dimensional reduction of
the fermions is that the charge of $\psi (x)$ is reversed as $x$ passes
through $0$.  ie. the conserved electric charge operator is:
\begin{equation} Q= 2\left[\Pi + \int_{-\infty}^{\infty}dx q(x)J^i_i
(x)\right] \label{Qdef}\end{equation} where the current operator is
\begin{equation} J^i_j (x) \equiv \psi^{\dagger
i}(x)\psi_j(x)\label{J}\end{equation} and  \begin{equation} dq/dx=-f
\hskip1cm \hbox{with} \hskip1cm q(x) \to \mp 1/2 \hskip1cm \hbox{as}
\hskip1cm x \to \pm \infty \end{equation} Thus $\psi(x)$ has charge $+1$ for
$x>r_0$ and charge $-1$ for $x<-r_0$.  [Note the factor of $2$ in Eq.
(\ref{Qdef}) which does not appear in Ref. (\onlinecite{Pol}). We insert it
so that the fermions have  charge $\pm 1$ rather than  $\pm 1/2$.] Although
$Q$ naively appears to commute with the Hamiltonian  it actually fails to do
so due to the anomalous Schwinger term in the current commutator:
\begin{equation} [J(x),J(y)] = -{iN\over 2\pi}\delta
^{'}(x-y)\label{Schwinger}\end{equation} Here, and henceforth,
\begin{equation} J \equiv J^i_i\end{equation} This problem can be corrected
by adding an additional term to the Hamiltonian: \begin{equation} H \to H +
{1\over 2}C\alpha^2\label{Hadd}\end{equation} where the constant, $C$, is
given by: \begin{equation} C \equiv {N\over 2\pi}\int_{-\infty}^{\infty}dx
f^2(x) \end{equation} Despite the fact that the part of $H$ involving
$\alpha$ only is of
 harmonic oscillator type, not invariant under translating $\alpha$, the full
Hamiltonian is invariant  under the combination of translation of $\alpha$
and phase rotation on the fermions induced by the operator $Q$.

Apart {}from the electric charge, $Q$ associated with a gauge symmetry, it is
also crucial to consider the baryon number density.  This is anomalous in the
standard electro-weak model and its grand-unified generalizations.  This
means that there does not exist a conserved gauge-invariant baryon number
current.  Instead we can define either a current, $J$ which is conserved but
not gauge-invariant or else a current $\tilde J$ which is gauge invariant
but not conserved.  It is the latter quantity which corresponds to the
observable baryon number and which changes during the process of
fermion-dyon scattering.  In the reduced model of Polchinksi, Eq.
(\ref{Ham}), (\ref{Hadd}), the conserved, but not gauge-invariant baryon
number density is simply the current operator $J(x)$ defined above in Eq.
(\ref{J}).  Including the anomly, \begin{equation} {\partial J\over \partial
t} = -{\partial J\over \partial x} +\alpha {N\over 2\pi} {df\over dx}
\end{equation} Thus, $\int_{-\infty}^{\infty}dx J(x)$ is time-independent,
assuming that $J(x)$ vanishes at spatial infinity.  On the other hand this
current does not commute with $Q$: \begin{equation} [J(x),Q] = {iN\over
\pi}f(x)\end{equation}   Temporarily reflecting the  incoming wave back to
the positive axis, we see that the electric charge density contains
$J_L-J_R$ whereas the baryon number density is $J_L+J_R$. The
non-commutativity of these operators is the cause of the anomaly in massless
(1+1)-dimensional Q.E.D.  This problem is solved by adding an extra term to
$J(x)$: \begin{equation} \tilde J(t,x) \equiv J(x) -{N\over 2\pi}\alpha
(t)f(x) \end{equation} giving a current which commutes with $Q$ (ie. is
``gauge invariant'') but not with $H$ (ie. is not conserved). Explicity,
\begin{equation} {\partial \tilde J \over \partial t}=- {\partial \tilde J
\over \partial x} -{Nf\over 2\pi I}\Pi \label{tilJeqm}\end{equation}

 The anomalous commutator of $\tilde J$ with itself is also given by Eq.
(\ref{Schwinger}), unaffected by the additional term.  Although the
motivation is somewhat different, there is a striking analogy between the
shift {}from $J$ to $\tilde J$ in the dyon problem and a similar shift which
is used in the conformal field theory analysis of the Kondo effect.\cite{A}

Polchinski has argued\cite{Pol} that at low energies, we may eliminate the
rotor degree of freedom and simply replace it by a boundary condition on the
fermions.   [This was  shown earlier by Callan and Das\cite{Call,Call-Das}
{}from a somewhat different perspective.]  It proves simplest to work with the
gauge-invariant current, $\tilde J$.  The equation of motion for $\Pi$ can be
expressed in terms of $\tilde J$ as: \begin{equation} {d \Pi \over
dt}=\int_{-\infty}^{\infty} dxf(x)\tilde J(x) \label{Pieqm}\end{equation} The
pair of linear equations, (\ref{tilJeqm}) and (\ref{Pieqm}) can be
integrated to find the time evolution of $\tilde J$ and $\Pi$ given their
initial values.  Integrating Eq. (\ref{tilJeqm}) gives: \begin{equation}
\tilde J(x,t) = \tilde J(-r_0, t-x-r_0)-{N\over 2\pi
I}\int_{-r_0}^xdx'f(x')\Pi (t+x'-x) \label{Jt}\end{equation} We are
interested in incoming particles of energy $E<<1/r_0 \approx 1/I$.  As we
shall see, this implies that $\tilde J(x,t)$ and $\Pi (t)$ vary slowly in
time (apart {}from a transient part of $\Pi$ which can be ignored).  Thus we
may approximate Eq. (\ref{Jt}) by: \begin{equation} \tilde J(x,t) \approx
\tilde J(-r_0,t)+{N\over 2\pi I}\Pi
(t)[q(x)-q(-r_0)]\label{Jsol}\end{equation} for $x$ of $O(r_0)$.  Plugging
this into Eq. (\ref{Pieqm}) gives: \begin{equation} {d\Pi \over dt}\approx
\tilde J(-r_0,t)-{N\over 4\pi I}\Pi \end{equation}  This gives:
\begin{equation} \Pi (t) = e^{-{N\over 4\pi I}t}\Pi (0) +
\int_0^tdt'e^{-{N\over 4\pi I}(t-t')}\tilde J(-r_0,t')\end{equation}
Assuming that $\tilde J(-r_0,t)$ varies slowly in time, and assuming
$t>>1/I$ so that the initial value of $\Pi (0)$ has decayed to essentially
$0$, we may approximate: \begin{equation} \Pi (t) \approx {4\pi I\over N}
\tilde J(-r_0,t) \label{Pisol}\end{equation} Thus we see that $\Pi (t)$ and
hence $\tilde J(x,t)$ vary slowly in time as assumed.  Finally plugging Eq.
( \ref{Pisol}) into Eq. (\ref{Jsol}) we obtain: \begin{equation} \tilde
J(r_0,t) \approx -\tilde J(-r_0,t)\label{Jbound}\end{equation} We see that
$\tilde J(x,t)$ flips sign as it passes by the dyon! This implies violation
of baryon number conservation.\cite{Call-Das,Pol} On the other hand, the
$SU(N)$ currents, \begin{equation} J^A \equiv \psi^{\dagger
i}(T^A)_i^j\psi_j\end{equation} with $trT^A=0$, and
$trT^AT^B=(1/2)\delta^{AB}$, do not couple to the rotor and suffer no
anomaly.  They are both conserved and gauge-invariant.  Consequently, they
vary slowly in space near the dyon: \begin{equation} J^A(r_0,t) \approx
J^A(-r_0,t)\label{JAbound}\end{equation}

It appears that it may be possible to ignore the dyon completely, at low
energies and long distances, and simply impose the boundary conditions of Eq.
(\ref{Jbound}) and (\ref{JAbound}) on the fermions. ie., we formally let
$r_0\to 0$ and require that  \begin{equation} \tilde J(0^+)=-\tilde
J(0^-)\end{equation} while requiring $J^A(x)$ to be continuous at the
origin. As was observed by Callan and Das\cite{Call-Das} and by
Polchinski\cite{Pol}, this only works straightforwardly for $N=1$ or $2$.
ie., only in these cases are the boundary conditions on the currents
equivalent to linear boundary conditions on the fermion fields.  For $N=1$
this is acheived by the condition: \begin{equation} \psi
(0^-)=e^{i\theta}\psi^{\dagger} (0^+)\end{equation} Here $\theta$ is an
arbitrary angle, which, as Polchinski argued, should be identified with the
topological angle of the non-abelian gauge theory.  For $N=2$ the boundary
conditions are: \begin{equation} \psi_i(0^-) = ie^{i\theta}
\epsilon_{ij}\psi^{\dagger j}(0^+)\label{N=2bc}\end{equation} Note that the
latter conserves the $SU(2)$ symmetry.

For $N>2$ no linear boundary condition on the fermions is equivalent to the
boundary conditions of Eq. (\ref{Jbound}) and (\ref{JAbound}) on the
currents.\cite{Call-Das,Pol}  This can be easily seen because such a linear
condition would have to take the form: \begin{equation}
\psi_i(0^-)=U_{ij}\psi^{\dagger j}(0^+)\end{equation} for some unitary
matrix, $U$.  While this automatically gives the correct condition on the
charge current, Eq. (\ref{Jbound}), in order for it to give the correct
condition on the flavour currents, Eq. (\ref{JAbound}), the matrix $U$ must
satisfy the condition: \begin{equation} U^{\dagger}T^AU = -\left( T^A\right)
^t = -\left( T^A\right)^* \end{equation} for all $A$.  This would imply that
the fundamental representation of $SU(N)$ is real which is untrue for $N >2$.
 Instead the problem was solved by other means.  Essentially, instead of
eliminating the dyon, Polchinski integrated out the fermions.  This could be
done exactly using an identity due to Schwinger for the fermion determinant.
Note that the boundary conditions on the currents (and, for that matter the
boundary conditions on the fermions for the $N=1$ or $2$ cases) are
time-independent and scale-invariant.  At low energies and long distances
compared to the dyon size and energy-level spacing, $1/I$, no scale enters
into the behaviour of the massless fermions.  We will show in the next
section that it is possible to solve the $N>2$ problem in the same way as
for $N\leq 2$, to wit by simply imposing conformally invariant boundary
conditions on  non-interacting fermions.  However, the notion of ``boundary
condition'' must be generalized somewhat, following Cardy.\cite{Cardy}

\section{Boundary States} In this section we briefly review
Cardy's\cite{Cardy} boundary conformal field theory and then deduce the
boundary states corresponding to the dyon, for arbitrary number of flavours,
$N$.  This determines all low-energy properties of the system including  the
low energy excitation spectrum for a dyon in a finite spherical box and the
Green's functions, which are discussed in the next section.

A system defined on the half-space $x \geq 0$ with a boundary condition at
$x=0$ can still be invariant under the infinite-dimensional subgroup of the
conformal group which leaves the boundary invariant.  ie., writing
$z=\tau+ix$, where $\tau$ is imaginary time, we require invariance under
analytic transformations $z \to w(z)$ such that $w(\tau )^*=w(\tau )$.  The
boundary condition is assumed to imply that no momentum flows across the
boundary, ie.  \begin{equation} T_L (0, \tau ) =  T_R(0,\tau
)\label{0mom}\end{equation} where $T_L$ and $T_R$ are the left and right
components of the energy-momentum tensor. Since $T_L(x,t) = T_L(t+x)$ and
$T_R(x,t)=T_R(t-x)$, this implies that we may regard $T_L$ as the analytic
continuation of $T_R$ to the negative $x$-axis; ie.: \begin{equation}
T_L(x,t) = T_R(-x,t).\label{LR} \end{equation} This formulation was implicit
in Section II.

Cardy's formalism is based on modular invariance; ie. on the possibility of
exchanging space and imaginary time in a (1+1) dimensional conformal field
theory with boundaries.  Thus, we consider a conformal field theory defined
on an interval of length $l$ with  (in general different) conformally
invariant boundary conditions at $x=0$ and $x=l$ which we denote generically
by $A$ and $B$.
 It is convenient to consider the system at finite temperature, $T$, so that
the fields are defined on an interval of length $\beta \equiv 1/T$ in the
imaginary time direction with periodic boundary conditions (in the case of
bosonic fields).  The Hamiltonian, including the boundary conditions is
denoted $H_{AB}^l$ and the corresponding partition function:
\begin{equation} Z_{AB} \equiv tr e^{-\beta H_{AB}^l}\end{equation} Using
Eq. (\ref{LR}), at $x=0$ and the equivalent condition at $x=l$, we may
regard the system as consisting of right-movers only on an interval of
length $2l$. Thus $Z_{AB}$ must be a sum of characters corresponding to the
various conformal towers in the right-moving sector of the theory. ie.,
\begin{equation} Z_{AB}= \sum_p n^p_{AB} \chi_p (e^{-\pi \beta
/l})\label{ZAB1}\end{equation} where $p$ labels the various conformal
towers, the multiplicities, $n^p_{AB}$ are non-negative integers and
\begin{equation} \chi_p(q) \equiv \sum_m d_p^m q^{x_p+m-c/24}
\label{chagen}\end{equation} where  $x_p$ is the scaling dimension of the
corresponding primary field and the $d_p^m$'s are non-negative integer
degeneracies. $c$ is the Virasoro central charge, the conformal anomaly
parameter. For convenience, we define: \begin{equation} q \equiv e^{-\pi
\beta /l}. \end{equation}
 Cardy made the fundamental observation that we should be able to reinterpret
$Z_{AB}$ by regarding the periodic direction as being space and the other
one as being time.  Now the Hamiltonian is defined on a periodic interval of
length $\beta$; we denote it as $H_P^{\beta}$. The imaginary time interval
is $l$.  However, it is no longer appropriate to take an operator trace
since the system is not periodic in the new ``time'' direction.  Rather, we
must consider matrix elements between some states $<A|$ and $|B>$.  ie.
\begin{equation} Z_{AB} = <A|e^{-lH_P^\beta}|B>\end{equation} The {\it
boundary states}, $|A>$ and $|B>$ correspond to the {\it boundary
conditions} in the original formulation.  The zero momentum condition of Eq.
(\ref{0mom}) implies that all consistent boundary states must obey:
\begin{equation}[T_L(x)-T_R(x)]|A>=0 \label{Ish}\end{equation} Thus, in
particular, expanding in eigenstates of $H_P^l$, all states will have equal
left and right energies: $x\equiv x_L+x_R=2x_R$.  In fact, Eq. (\ref{Ish})
implies that all boundary states must contain infinite sums over all
descendents of a given primary state of the form: \begin{equation}
|p>=\sum_m|p;m>_L \otimes |p;m>_R\label{Ishsol}\end{equation} Here the
integer $m$ schematically labels all descendents of the primary states.  The
primary state $|p;0>$ has unit normalization.  The left and right conformal
towers occuring must be equivalent. Such states, corresponding to a
particular conformal tower, are known as Ishibashi states.\cite{Ishibashi}
Thus we again obtain the characters of the conformal towers but this time
$e^{(-4\pi l/\beta )x}$ occurs.  Defining  \begin{equation} \tilde q \equiv
e^{-4\pi l/ \beta }, \end{equation} we obtain: \begin{equation} Z_{AB} =
\sum_p<A|p><p|B> \chi_p(\tilde q)\label{ZAB2}\end{equation} Equating the
expressions of Eq. (\ref{ZAB1}) and (\ref{ZAB2}) gives a set of powerful
consistency conditions on possible boundary states $|A>$ and spectra,
$n_{AB}$, that we refer to as the Cardy conditions.  In principle, one
wishes to find a complete set of solutions of these equations in order to
enumerate all possible conformally invariant boundary conditions, or
equivalently boundary states, for a given problem, ie. a given $H^P$.  Given
any two consistent boundary states $|A>$ and $|B>$ the linear combination
$n|A>+m|B>$ where $n$ and $m$ are non-negative integers is always another
consistent state. Note that the partition function involving this state and
another one, $C$ is: \begin{equation} Z_{nA+mB,C} =
nZ_{AC}+mZ_{BC}\end{equation}
 We are interested in finding a complete basis of states {}from which all
solutions can be constructed in this way.  In the case of the Ising model,
for example, there are only three such basis boundary states corresponding
to spin-up, spin-down or free boundary conditions on the Ising spins.  Once
we have found the boundary states, we can directly obtain not only the
finite-size spectrum but also the Green's functions as is explained in the
next section.

We now consider the boundary states corresponding to the dyon. The periodic
Hamiltonian, $H^P$, for our problem is simply the Hamiltonian for $N$
flavours of free fermions (left and right movers).  We will also find the
boundary state corresponding to a trivial boundary condition on the
fermions.  For instance, we might wish to impose a vanishing boundary
condition in the three-dimensional problem on the surface of a sphere of
radius $l$, with the dyon at the centre.  In principle there could be many
possible boundary states for such a $c=N$ theory for large $N$.  Our search
is considerably simplified by the boundary conditions on the currents of Eq.
(\ref{Jbound})  and (\ref{JAbound}).  As Cardy argued, because the currents
have conformal spin $1$ these conditions pick up a relative minus sign under
the modular transformation.  Thus the flavour current boundary condition
becomes: \begin{equation} [J^E_L(x)+J^E_R(x)]|A>=0 \end{equation} (We hope
the reader is not confused by our notation.  The superscript $E$ labels the
generators of $SU(N)$.)  As remarked above, this condition holds both for
the free boundary condition and also for the dyon.  Likewise we have
conditions on the boundary states coming {}from the charge current boundary
condition.  This condition has a different form for the free and dyon
boundary condition: \begin{equation} [J_L(x) \pm
J_R(x)]|A>=0\label{JIshcond}\end{equation} where the upper and lower sign
refers to the free and dyon case respectively.  This change in sign for the
dyon case follows {}from Eq. (\ref{Jbound}).  These conditions allow us to
enormously reduce the set of Ishibashi states {}from which we build our
boundary states.

To take advantage of this simplification we use non-abelian
bosonization.\cite{NAB,KZ}  This means that we represent the $N$ flavours of
free fermions in terms of a level $k=1$ $SU(N)$ Wess-Zumino-Witten field, $g$
containing the flavour degrees of freedom together with a free boson, $\phi$,
representing the charge degrees of freedom.  The free fermion energy momentum
tensor for right-movers can be written entirely in terms of the flavour and
charge currents as: \begin{equation} T_R = {\pi \over N}J_RJ_R + {2\pi \over
N+1}J_R^AJ_R^A\end{equation} Here,
 \begin{equation} J_R=\sqrt{N/\pi}\partial \phi /\partial x \end{equation}
and $J^A_R$ can be expressed in terms of $g$. This already suggests a
simplification in our treatment of the dyon problem because the entire
interacting Hamiltonian  can be written as: \begin{equation} H =
\int_{-\infty}^{\infty} dx[{\pi \over N}J_RJ_R + {2\pi \over
N+1}J_R^AJ_R^A-\alpha f J_R]+ {\Pi^2\over 2I} + {C\over 2}\alpha ^2
\end{equation} Note that the interaction only involves the charge current.
The flavour degrees of freedom appear to play a completely passive role.
This is the opposite of what happens in the Kondo problem where it is the
spin currents which interact with the impurity and the charge current
remains non-interacting.\cite{A}

Let us first consider the properties of $H_P^\beta$ in more detail.  Because
$tre^{-\beta H_{AB}^l}$ for fermions involves {\it antiperiodic} boundary
conditions in the imaginary time direction, in order to maintain modular
invariance, we must also impose anti-periodic boundary conditions in the
space direction, in $H_P^\beta$.  [Actually, only one generator of the
modular group is used; switching space and imaginary time.]  The eigenstates
of $H_P^\beta$ then consists of a direct product of left and right
eigenstates. The allowed momenta are: \begin{equation} k=\pm (2\pi /\beta
)(n+1/2)\end{equation} where $n$ is a positive integer and the + or - sign
occurs for right-movers or left-movers respectively. The spectrum of
$H_P^\beta$ for one flavour of right-movers can be seen to be $E=(2\pi
/\beta )x$ where \begin{equation} x=-1/24+Q^2/ 2+\sum_{m=1}^\infty
n_{m}m\label{spec-per}\end{equation} Here $Q$, an integer, is the
fermion-number of the state and thd $n_m$'s are non-negative integers. We
have included the universal $O(1/\beta )$ term in the groundstate
energy.\cite{c}  A derivation of this formula is given in Ref.
(\onlinecite{A}).  It can be understood {}from abelian bosonization.  We write
the right-moving fermion in terms of a right-moving boson: \begin{equation}
\psi_R \propto e^{i\sqrt{4\pi}\phi_R}\end{equation} Taking into account that
$\phi_R$ does not commute with itself at different spatial points we
conclude that {\it periodic} boundary conditions should be imposed on it.
It then has the mode expansion: \begin{equation}
\phi_R(x,t)=\phi_{0R}+{\sqrt{\pi}Q(x-t)\over \beta} + \sum_{m=1}^\infty
{1\over \sqrt{4\pi m}}\left(e^{-2\pi im(t-x)/\beta}a_{mR}+
h.c.\right)\label{mode-exp}\end{equation} where $Q$ is an integer, the
soliton number.  Plugging this mode expansion into the Hamiltonian:
\begin{equation} H_R=(1/2)\int_0^\beta dx[(\partial \phi_R/\partial t)^2+
(\partial \phi_R/\partial x)^2]\end{equation} we obtain the spectrum of Eq.
(\ref{spec-per}).   $n_{m}$ labels the occupation number of the
$m^{\hbox{th}}$ harmonic mode, ie. the eigenvalue of $a_m^\dagger a_m$.
With $N$ flavours we may write the spectrum as $N$ copies of the above.
Alternatively, using non-abelian bosonization, we separate it into a charge
and flavour part. [See, for example Ref. (\onlinecite{A}).] The charge
boson, $\phi$ can be regarded as the canonically normalized sum of bosons
for each flavour: \begin{equation} \phi_R \equiv
(1/\sqrt{N})\sum_{i=1}^N\phi_R^i\end{equation} It has an identical mode
expansion to Eq. (\ref{mode-exp}) except that $Q$ is replaced by
$Q/\sqrt{N}$.  Now $Q$ has the interpretation of the total fermion number.
Consequently, the spectrum of right-moving charge excitations is:
\begin{equation} x=-1/24+Q^2/ 2N+\sum_{m=1}^\infty n_{m}m\end{equation} If
we classify states into conformal towers using the $U(1)$ current algebra as
well as the Virasoro algebra, then we obtain one conformal tower for each
integer value of $Q$, all states in a given conformal tower having the same
charge, $Q$. The corresponding character is: \begin{equation} \chi_Q (q)
\equiv \sum_{n_1=0}^\infty \sum_{n_2=0}^\infty ...q^{-1/24+Q^2/
2N+\sum_{m=1}^\infty n_{m}m}=[\eta (q)]^{-1}q^{Q^2/2N} \end{equation} where
\begin{equation} \eta (q) \equiv q^{1/24}\prod_{m=1}^\infty
(1-q^m)\end{equation}

There are $N$ different conformal towers of $SU(N)_1$;\cite{KZ} the highest
weight states transform under the antisymmetric $p$-index  tensor
representation of $SU(N)$ (ie. the representation whose Young tableau
consists of a single column of length $p$), for $p=0,1,... N-1$.  The
corresponding scaled energies are: \begin{equation} x_p =
-(N-1)/24+p(N-p)/2N \end{equation} Certain ``gluing conditions'' determine
which combinations of flavour and charge conformal towers occur in the
spectrum with periodic boundary conditions.  These can be shown to
be:\cite{A} \begin{equation} Q=p\ (\hbox{mod}\  N) \end{equation} [Recall
that $p$ labels the global $SU(N)$ representation of the highest weight
state in the conformal tower; descendents can transform under other
representations of global $SU(N)$.]  Note, for example, that the lowest
energy excitation has $p=Q=1$.  It has $x=(N-1)/2N + 1/2N = 1/2$ and
corresponds to a single fermion state which does indeed have these quantum
numbers and energy [fundamental representation, charge $1$, energy $(\pi
/\beta$)].  Clearly a state with $Q=1$, $p=0$, with $x= 1/2N$ should not be
permitted since it does not correspond to any free fermion state.

The above discussion was given for the right-moving sector.  The left-moving
sector has an identical spectrum with corresponding conformal towers
labelled by
 $Q_L$ and  $p_L$.   The gluing conditions are imposed separately on left
and right sectors: \begin{eqnarray} Q_R&=&p_R \ (\hbox{mod}\ N)\nonumber \\
Q_L&=&p_L \ (\hbox{mod}\ N)\end{eqnarray}

Let us now consider the relevant Ishibashi states. We first discuss the
charge states in some detail.  Considering first the condition
$[J_L(x)-J_R(x)]|A>=0$, this clearly implies $Q_L=Q_R$.  Furthermore, when
we sum over the occupation number of all finite-momentum harmonic modes,
created by the operators, $a_{L m}^{\dagger}$ and $a_{R m}^{\dagger}$ where
$m$ labels momentum, we must ensure that all corresponding left and right
occupation numbers are equal. This means that the  Ishibashi states take the
form: \begin{equation} |Q>_{c -} \equiv  |Q>_R\otimes |Q>_L\otimes
\exp[\sum_{m=1}^\infty a_{R m}^{\dagger}a_{L
 m}^{\dagger}]|0>\end{equation} The first two kets represent the soliton
modes of the right and left moving bosons respectively and $|0>$ represents
the direct product of vacuum states for left and right harmonic modes.  The
other set of Ishibashi states obeying the condition $[J_L(x)+J_R(x)]|A>=0$
can be obtained {}from this one by simply making a unitary transformation that
flips the sign of $J_L$ while leaving $J_R$ invariant.  This corresponds to
taking $Q_L \to -Q_L$ and $a_{L m}^\dagger \to -a_{L m}^\dagger$.  This
state is: \begin{equation} |Q>_{c +} \equiv  |Q>_R\otimes |-Q>_L
\exp[-\sum_{m=1}^\infty a_{R m}^{\dagger}a_{L
 m}^{\dagger}]|0>\label{Qc+}\end{equation} The Ishibashi states in the
flavour sector take the form: \begin{equation} |p>_f \equiv
\sum_{m=1}^{\infty}|p,m>_R\otimes U|p,m>_L. \end{equation} Here $p$ labels
the $N-1$ flavour conformal towers and $m$ generically labels all
descendents.  The anti-unitary operator $U$ has the property $J_L^A U =
-UJ_L^A$. In particular, it takes $p_L \to N-p_L$.

 All boundary states that we need will be linear combinations of products of
the form $|Q>_{c \pm} \otimes |p>_f$.  An immediate restriction on the
possibilities arises {}from the gluing conditions inherent in $H^P$, namely
$Q_L=p_L\ (\hbox{mod}\  N)$, $Q_R=p_R\ (\hbox{mod}\  N)$.  The only products
of charge and flavour Ishibashi states involving $|Q>_{c +}$, which satisfy
these conditions obey $Q=p\ (\hbox{mod}\  N)$.  Note that $|Q>_{c +}$
contains charge $Q_L=-Q_R=-Q$ and $|p>_f$ contains the states in the
conformal tower $p_L=N-p_R=N-p$, satisfying the gluing condition.  If we use
the charge states $|Q>_{c -}$, the possiblities are even more restricted.
Since $|Q>_{c -}$ contains only states with $Q_L=Q_R=Q$ while $|p>_f$
contains only states in the conformal towers with $p_R=p$, $p_L=N-p$ we see
that there is only one way of satisfying the gluing conditions for odd $N$;
we must choose $p=0$, $Q=0 \  (\hbox{mod}\  N)$.  In the case of even $N$
there is one other possibility; $p=N/2$, $Q=N/2\ (\hbox{mod}\  N)$ since
then $Q_L=N/2\ (\hbox{mod}\  N)$, $Q_R=N/2\ (\hbox{mod}\  N)$ and
$p_L=p_R=N/2$.  The gluing conditions inherent in $H^P$ and the boundary
conditions on the currents have restricted the possible boundary states to
such an extent that it is now straightforward to construct the needed
boundary states corresponding to free fermions or a dyon.

We begin with the case of free fermions.  A convenient set of conformally
invariant boundary conditions is: \begin{equation} \psi_{Ri}(0) =
e^{-i\theta}\psi_{Li}(0) \label{freebc}\end{equation} Thus we may
analytically continue to $x<0$, defining: \begin{equation} \psi_{Ri}(-x,t)
\equiv  e^{-i\theta}\psi_{Li}(x,t)\end{equation} If we also impose the
boundary condition at $l$: \begin{equation} \psi_{Ri}(l) =
-e^{-i\theta^{'}}\psi_{Li}(l),\label{freebc2}\end{equation} then the problem
is equivalent to having right-movers only on a circle of circumference $2l$
with the twisted boundary condition: \begin{equation}
\psi_{Ri}(-l)=-e^{i(\theta '-\theta )}\psi_{Ri}(l) \end{equation} The
parameter $\theta$ corresponds to a phase shift. The allowed momenta of the
fermions are  \begin{equation} k=(\pi /l)[ n + 1/2\pm (\theta -\theta '
)/2\pi ] \end{equation} where the upper (lower) sign is for particles
(anti-particles) and $n$ must be an integer such that $k\geq 0$.   The
spectrum can be written as: [See ref. (\onlinecite{A}).] \begin{equation} E
= {\pi \over l}\left\{-N/24+{1\over 2N}\left[Q+{(\theta '-\theta )\over
2\pi}N\right]^2+{p(N-p)\over 2N} + n_c + n_f\right\}\end{equation} Here
$n_c$ and $n_f$ are integers labelling descendents in the charge and flavour
sector respectively; $Q$ and $p$ label the charge and flavour conformal
towers to which the state belongs. The gluing condition $Q=p \ (\hbox{mod}\
N)$ must be imposed. We have included the $O(1/l)$ contribution to the
groundstate energy.  The corresponding partition function is:
\begin{equation} Z_{F\theta ,F\theta^{'}} =
\sum_{p=0}^{N-1}\chi_p^f(q)\sum_{m=-\infty}^\infty \chi^c_{p+mN}(q,\theta '
-\theta) \label{ZFF1}\end{equation} Here $\chi^f_p(q)$ is the character of
the $p^{th}$ flavour conformal tower.  $\chi^c_Q (q,\theta )$ is the
character of the charge $Q$ conformal tower with the phase shift $\theta$
included.  Explicitly: \begin{equation} \chi_Q^c(q,\theta ) \equiv {1\over
\eta (q)}q^{(Q+\theta N/2\pi )^2/2N},\end{equation} Note that
\begin{equation}\chi_Q^c(q,\theta +2\pi /N)=\chi_{Q+1}^c(q,\theta )
\label{chicthe}\end{equation}
 so a relative phase shift of $2\pi /N$  is equivalent to shifting the gluing
conditions to $Q=p+1 (\hbox{mod}\  N)$.  This observation is basic to the
conformal field theory treatment of the Kondo problem.\cite{A}

We wish to find the corresponding boundary states $|F\theta >$, such that:
\begin{equation} Z_{F\theta ,F\theta^{'}}=<F\theta
|e^{-lH_P^\beta}|F\theta^{'}> \label{ZFF2}\end{equation} Since Eq.
(\ref{ZFF2}) can be expressed as a sum over conformal dimensions, $x$ of
$\tilde q^x$, it is convenient to express $Z_{F\theta ,F\theta^{'}}$ in Eq.
(\ref{ZFF1}) in terms of $\tilde q$.  The needed modular transformation
involves the so-called modular S-matrix.  We work this out explicitly in the
charge case.  The modular transformation of $\eta$ is: \begin{equation} \eta
(q)=\sqrt{2l/\beta}\eta (\tilde q)\end{equation} [See for example Ref.
(\onlinecite{Gins})]. While $\chi^c_Q(q,\theta )$ does not itself have nice
modular transformation properties, it turns out that the sum of this
quantity over $Q$ in Eq. (\ref{ZFF1}) does.  We use the Poisson sum formula,
ie. the Fourier transform of the periodic $\delta$-function, $\delta_P(x)$:
\begin{equation} \delta_P(x)\equiv \sum_{m=-\infty}^\infty \delta (x-m) =
\sum_{Q=-\infty}^\infty e^{2\pi iQx}\end{equation}This gives:
\begin{equation} \sum_{m=-\infty}^\infty q^{(p+mN+\theta N/2\pi
)^2/2N}=\sqrt{2l/\beta N}\sum_{Q=-\infty}^\infty e^{-iQ(2\pi p/N+\theta
)}\tilde q^{Q^2/2N} \end{equation} Hence, \begin{equation}
\sum_{m=-\infty}^\infty \chi^c_{p+mN}(q,\theta )={1\over \sqrt{N}}
\sum_{Q=-\infty}^\infty e^{-i(2\pi p/N+\theta )Q}\chi^c_Q(\tilde
q,0)\label{cmod}\end{equation} We also need the modular transformation
property of the flavour characters.  (The explicit values of these
characters is not needed.)  It is found that,\cite{ABI} \begin{equation}
\chi_p^f(q)=\sum_{p'=0}^{N-1}{1\over \sqrt{N}}e^{2\pi
ipp'/N}\chi_{p'}^f(\tilde q)\label{fmod}\end{equation} We now substitute Eq.
(\ref{cmod}) and (\ref{fmod}) into Eq. (\ref{ZFF1}).  Using the identity:
\begin{equation} {1\over N}\sum_{p=0}^{N-1}e^{2\pi
ip(p'-Q)/N}=\sum_{m'=-\infty}^\infty \delta_{Q,m'N+p'},\end{equation} we
obtain finally: \begin{equation} Z_{F\theta ,F \theta '}=\sum_{p=0}^{N-1}
\sum_{m=-\infty}^\infty  e^{i(\theta '-\theta )(mN+p)} \chi^f_p(\tilde q)
\chi^c_{p+mN}(\tilde q,0)\label{ZFF} \end{equation} We now look for the
boundary states $|F \theta >$ using Eq. (\ref{ZAB2}). Note that we must use
the charge Ishibashi states $|Q>_{c +}$ of Eq. (\ref{Qc+}) in this case in
order to satisfy the condition of Eq. (\ref{JIshcond}) on the boundary
states with the positive sign.  Let us write the state as: \begin{equation}
|F\theta > = \sum_{p=0}^{N-1} \sum_{m=-\infty}^\infty a_{mN+p}(\theta
)|mN+p>_{c+}\otimes |p>_f \label{Fstate1}\end{equation} Let us begin with
the case $\theta =0$.  It can easily be shown that the products of
characters, $\chi^f_p(\tilde q) \chi^c_{p+mN}(\tilde q)$ are linearly
independent for different values of $p, N$, apart {}from the equality:
\begin{equation}\chi^f_p(\tilde q) \chi^c_{p+mN}(\tilde
q)=\chi^f_{N-p}(\tilde q) \chi^c_{-p-mN}(\tilde q)\end{equation} Thus, {}from
considering $Z_{F 0,F 0}$ we see that: \begin{equation}
|a_Q(0)|^2+|a_{-Q}(0)|^2=2,\end{equation} for all $Q$.  Since the boundary
condition of Eq. (\ref{freebc}), (\ref{freebc2}) is charge conjugation
invariant,for $\theta =0$, we should impose this symmetry on the
corresponding state, $|F 0>$.  Thus, $a_Q=a_{-Q}$. Thus we obtain
$|a_Q(0)|=1$, for all $Q$.  We may then redefine the phases of the Ishibashi
states so that: \begin{equation} a_Q(0)=1\end{equation}

We pause to comment on the significance of charge conjugation in the
underlying three-dimensional theory.  The symmetry which we are interested
in changes the sign of the electric charge, but not the magnetic charge of
the dyon.  This symmetry is CP.  Note that parity, P, {\it does not} take $x
\to -x$ in the dimensionally reduced theory.  The reason is that we are
dealing with s-wave components of three-dimensional fields; $x$ actually
corresponds to the radial variable, $r$.  ($x$ can be negative as well as
positive because we have reflected the incoming wave to the negative axis.)
Thus three-dimensional CP
 corresponds to one-dimensional C.  We shall hencforth refer to this
symmetry by its three-dimensional name, CP.

A simple solution, consistent with Eq. (\ref{ZFF}), for general $\theta$ is:
\begin{equation} a_Q(\theta )=e^{iQ\theta}\label{Fstate2}\end{equation} This
can be seen to be the unique solution which reproduces all the Green's
functions correctly, by the procedure explained in the next section.  We
note that it can be obtained {}from the $\theta =0$ state by a unitary
transformation: \begin{equation} |F\theta >=e^{i\hat Q\theta
/2}|F0>\label{unitary}\end{equation} where $\hat Q$ is the charge operator
defined in Eq. (\ref{Qdef}).

We now wish to determine the boundary state, $|D>$,  corresponding to the
dyon.  In this case, we don't know, a priori, the spectrum so we must
proceed somewhat differently.  In fact we will find a one-parameter family
of solutions, analogous to the free fermion states, $|F \theta>$ found
above.  We will argue that this parameter should be identified with the
gauge theory topological angle.  Fortunately, the arguments given above
drastically restrict the possible form of $|D>$. The most general possible
form which is equally valid for  $N$ even or odd is:  \begin{equation} |D> =
\sum_{m=-\infty}^\infty a_m|mN>_{c -}\otimes |0>_f
\label{Dodd}\end{equation} where the $a_m$ are free parameters. Note that we
must use the $|Q>_{c-}$ states in this case to satisfy the reversed sign
charge current boundary condition of Eq. (\ref{JIshcond}).
 We restrict the $a_m$'s by considering the  matrix element
$<D|e^{-lH_P^\beta }|D>$.  This formally corresponds to having  ``dyon
boundary conditions'' at both ends of the line interval.  It is difficult to
imagine a physical situation in three dimensions that  corresponds to this,
but it is certainly possible in the reduced one-dimensional theory.  We
insist, following Cardy, that this matrix element correspond to a
well-behaved partition function; ie. it should have the general form of Eq.
(\ref{ZAB1}), (\ref{chagen}) for some {\it integers} $d_p^m$ and
$n^p_{AB}$.  The matrix element is given by: \begin{equation}
Z_{DD}=\chi_0^f(\tilde q)\sum_{m=0}^\infty |a_m|^2\chi^c_{mN}(\tilde
q)\end{equation} In order to ensure that this gives a well-behaved partition
function, we need to modular transform.  However, the modular transform of a
single charge character, $\chi_Q(\tilde q)$ involves an {\it integral} over
a continuous range of charge conformal towers.  ie.: \begin{equation}
{1\over \eta (\tilde q)}\tilde q^{m^2N/2}={1\over \sqrt{N}\eta
(q)}\int_{-\infty}^\infty dx q^{x^2/2N}e^{2\pi imx}\end{equation} Thus, we
obtain: \begin{equation} Z_{DD}={1\over N}\sum_{p=0}^{N-1}\chi^f_p(q){1\over
\eta (q)}\int_{-\infty}^\infty dx q^{x^2/2N}\sum_{m=-\infty}^\infty  |a_m|^2
e^{2\pi imx}\end{equation} At this point we also require CP invariance.
This holds provided that the topological angle is $0$ or  $\pi$ in the
four-dimensional non-abelian gauge theory.  Hence, $a_m=a_{-m}$.  If we also
require that the groundstate be unique, then: \begin{equation}
\sum_{m=-\infty}^\infty |a_m|^2 e^{2\pi imx}=N\delta_P(x)
\label{amo}\end{equation}
  The spectrum is then: \begin{equation}
Z_{DD}=\sum_{p=0}^{N-1}\chi^f_p(q)\chi^c_Q(q ,0)\end{equation} Eq.
(\ref{amo}) then determines $|a_m|^2=N$.  We may choose the phase of the
$|Q>_{c-}$ Ishibashi states so that $a_m=\sqrt{N}$.

We should consider the matrix elements between free and dyon states.  This
corresponds to a situation where a dyon, corresponding to boundary state
$|D>$, given in Eq. (\ref{Dodd}), is at the centre of a spherical box of
radius $l$.  At the surface of the box we impose the free boundary condition
with phase shift $\theta$: \begin{equation} \psi_R(l)=-e^{-i\theta
}\psi_L(l)\end{equation} We need the matrix element $_{c +}<Q|e^{-lH_P^\beta
}|Q'>_{c -}$.  We see {}from the definitions that this vanishes unless
$Q=Q'=0$.   In this case the overlap of the coherent states involving the
harmonic modes gives: \begin{equation}  _{c +}<0|e^{-lH_P^\beta }|0>_{c -} =
\tilde  q^{-1/24}\prod_{n=1}^\infty (1+\tilde q^n)^{-1}\equiv W(\tilde q)
\end{equation} To proceed, we need the modular transform of $W(\tilde q)$.
This can be obtained {}from the Euler identity, the Jacobi triple product
formula and Eq. (\ref{cmod}): \begin{equation}  W(\tilde q) = \sqrt{2}W_+(
q) \equiv {1\over \eta ( q)\sqrt{2}}\sum_{Q=-\infty}^\infty
q^{(Q+1/2)^2/4}\end{equation}[See, for example, Ref. (\onlinecite{Gins}).]
Also modular transforming the $\chi_p^f$, using Eq. (\ref{fmod}), we obtain:
\begin{equation} Z_{F \theta ,D}\equiv <F \theta |e^{-lH_P^\beta } |D
>=\sum_{p=0}^{N-1}\chi^f_p(q)a_0\sqrt{2/N}W_+(q)\label{ZFD}\end{equation}
We see that we must require  $a_0=\sqrt{N/2}$ to obtain a unique
groundstate.  But {}from considering $Z_{DD}$, above, we concluded that
$a_0=\sqrt{N}$. It is impossible to find values for the $a_m$'s such that
the groundstate is non-degenerate in both $Z_{DD}$ and $Z_{F\theta ,D}$. We
see that the solution with the minimal possible groundstate degeneracy in
$Z_{DD}$ and $Z_{F\theta ,D}$, is $a_0=\sqrt{2N}$, giving a two-fold
degenerate groundstate in both partition functions.

For even $N$ a more general boundary state of dyon type exists.  We will find
that the peculiar degeneracy mentioned above can be avoided in this case.
Indeed, the odd $N$ problem does not arise {}from a well-defined
(3+1)-dimensional field theory and is believed to be
pathological.\cite{Witten2}  Polchinski\cite{Pol} showed that the Green's
functions exhibit unphysical behaviour for odd $N$.  The most general
solution for even $N$, consistent with the current boundary conditions and
the gluing conditions is: \begin{equation} |D> = \sum_{m=-\infty}^\infty
[a_m|mN>_{c -}\otimes |0>_f+b_m|(m+1/2)N>_{c -}\otimes |N/2>_f
]\label{Deven}\end{equation} Modular transforming we obtain:
\begin{equation} Z_{DD}={1\over N}\sum_{p=0}^{N-1}\chi^f_p(q){1\over \eta
(q)}\int_{-\infty}^\infty dx q^{x^2/2N}\sum_{m=-\infty}^\infty  [|a_m|^2
e^{2\pi imx}+(-1)^p|b_m|^2e^{2\pi i(m+1/2)x}]\end{equation}  Now the choice,
$a_m=b_m=\sqrt{N/2}$ gives the partition function: \begin{equation}
Z_{DD'}=\sum_{p=0}^{N-1}\chi^f_p(q)\sum_{Q=-\infty}^{\infty} {1\over
2}[1+(-1)^{p+Q}]\chi^c_Q(q,0)\label{ZDD}\end{equation} This is the unique
choice giving a non-degenerate, CP invariant groundstate, using  our phase
freedom to make all the $a_m$'s and $b_m$'s real and positive. The dyon-free
fermion partition function is again given by Eq. (\ref{ZFD}) with
$a_0=\sqrt{N/2}$.  In this case, we have  well-behaved partition functions
with unique groundstates in both dyon-dyon and dyon-free cases.  We refer to
this CP invariant state as $|D0>$.

Similar to the free case, we may find a simple one-parameter generalization
of this state by making a unitary transformation: \begin{equation} |D\theta
>\equiv e^{i\hat B\theta /N}|D0>\end{equation} Here \begin{equation}\hat
B\equiv \int_{-\infty}^\infty dx\tilde J(x)\label{unitdyon}\end{equation} is
the baryon number.  [Note that we must use the baryon number, $\hat B$, here
rather than the electric charge, $\hat Q$ as in Eq. (\ref{unitary}) since
the $|Q>_{c+}$ states occur in this case.]  The unitary transformation of
Eq. (\ref{unitdyon}) is precisely the one that connects the various
``$\theta$-vacua'' in the (3+1)-dimensional gauge theory. ie. the full
Hilbert Space separates into sectors labelled by the angle, $\theta$.
Making this unitary transformation is equivalent to introducing the
topological angle into the Lagrangian. [See, for example the discussion in
Ref. (\onlinecite{Pol}) where $B$ is referred to as $Q_5$, the chiral
charge.]  Thus we may identify the CP non-invariant state, $|D\theta >$, as
the dyon boundary state in the theory with topological angle, $\theta$.  We
will confirm this in the next section by computing the Green's functions and
comparing to the results of Polchinski\cite{Pol}.
 The corresponding coefficients, defined in Eq. (\ref{Deven}) are:
 \begin{equation} a_m = \sqrt{N\over 2}e^{im2\theta}, \ \  b_m = \sqrt{N\over
2}e^{i(2m+1)\theta}\label{abeven}\end{equation} Explicitly, \begin{equation}
|D\theta>\equiv \sqrt{N\over 2}\sum_{m=-\infty}^\infty
\left[e^{i2m\theta}|mN>_{c-}\otimes |0>_f+e^{i(2m+1)\theta}|(m+1/2)N>_{c-}
\otimes |N/2>_f\right]\label{Dtheta}\end{equation} The dyon boundary state,
$|D\theta >$ of Eq. (\ref{Dtheta}), is the main result of this paper.

Note that the linear combination of  boundary states: \begin{equation} |D
\theta > + |D \theta +\pi > =\sqrt{2N} \sum_{m=-\infty}^\infty
e^{2im\theta}|mN>_{c -}\otimes |0>_f,\end{equation} has the form of Eq.
(\ref{Dodd}). In other words, for even $N$, that state is actually a linear
combination of two basis boundary states, which explains the factor of two
degeneracy.  On the other hand, for odd $N$, it cannot be written as a
linear combination and the degeneracy cannot be eliminated.

The corresponding spectrum  for two states $|D\theta >$ and $|D\theta '>$
is:\begin{eqnarray}Z_{D\theta ,D\theta'}
&=&\sum_{p=0}^{N-1}\chi^f_p(q)\sum_{Q=-\infty}^\infty \chi^c_Q[q,2(\theta
'-\theta )/N]{1\over 2}[1+(-1)^{p+Q}]\nonumber \\
&=&\sum_{p=0}^{N-1}\chi^f_p(q)\sum_{Q=p \ (\hbox{mod}\ \
2)}\chi^c_Q[q,2(\theta ' -\theta )/N]\end{eqnarray}  Note that the
restriction on combining charge and flavour excitations, $p=Q\ (\hbox{mod}\
2)$ is  less restrictive than the gluing condition in the free case, $p=Q\
(\hbox{mod}\  N)$ for even $N\geq 4$.  It leads to the occurance of states
with exotic quantum numbers.  It can be shown\cite{Cardy} by a conformal
transformation that the primary states in the spectrum with $D\theta-D\theta
$ boundary conditions are in one-to-one correspondance with the boundary
operators present with a dyon boundary condition.  Hence these also have
exotic quantum numbers and scaling dimension, as we shall see explicitly in
the next section.  Note also that shifting $\theta$ by $\theta \to \theta
+\pi$ is equivalent to changing the gluing conditions to $p=Q+1\
(\hbox{mod}\  2)$ by Eq. (\ref{chicthe}).

The matrix element between the dyon state of Eq. (\ref{Deven}) and
(\ref{abeven}) and the free fermion state of Eq. (\ref{Fstate1}) and
(\ref{Fstate2}) is given by: \begin{equation}Z_{F\theta D\theta '}\equiv
<F\theta |e^{-lH^\beta_P}|D\theta '>=\sqrt{N\over 2}\chi^f_0(\tilde
q)W(\tilde q)=\sum_{p=0}^{N-1}\chi^f_p(q)W_+(q) \end{equation}
 This is again a consistent partition function with a non-degenerate
groundstate. This partition function for ``a dyon in a box'' has not, to our
knowledge, been obtained before.

The occurance of $W_+(q)$ can be understood by considering the effect of
imposing the charge boundary conditions: \begin{equation} J_R=\pm
J_L\label{Jcondsp}\end{equation} with opposite sign at the two ends of the
system.  These conditions essentially imply:
\begin{eqnarray}\phi_R(0)&=&\phi_L(0) \nonumber \\ \phi_R(l)&=&-\phi_L(l)
\end{eqnarray} Equivalently, working with a right-mover only:
\begin{equation} \phi_R(l)=-\phi_R(-l)\end{equation} This twisted boundary
condition determines the mode expansion: \begin{equation}
\phi_R(x-t)=\sum_{m=0}^\infty {1\over \sqrt{4\pi (m+1/2)}} \left[e^{-i2\pi
(m+1/2)(t-x)/2l}a_{m+1/2}+ h.c.\right] \end{equation} Note that unlike the
periodic case, there are no soliton modes and the harmonic modes have
half-integer, rather than integer frequencies.  The spectrum is now $E=(\pi
/l)x$ with \begin{equation} x = 1/48+ \sum_{m=0}^\infty
(m+1/2)n_{m+1/2}\end{equation} The partition function is thus:
\begin{equation} Z_{\hbox{twisted}}(q)=q^{1/48}\prod_{m=0}^\infty
[1-q^{m+1/2}]^{-1}\end{equation} {}from Euler's identity and the Jacobi triple
product formula we find: \begin{equation}
Z_{\hbox{twisted}}(q)=W_+(q)\end{equation}

On the other hand, the dyon-dyon partition function, like the free-free one,
involves the charge characters, $\chi_Q^c(q,\theta )$ rather than $W(q)$.
This follows because with the charge boundary conditions of Eq.
(\ref{Jcondsp}) with the same sign at the two ends, the boundary condition
in the  right-moving formulation is simply periodic: \begin{equation}
\phi_R(l)=\phi_R(-l)\ \  (\hbox{mod}\ \sqrt{\pi}) \end{equation}

To summarize the results of this section, we have studied the most general
conformally invariant boundary states consistent with the boundary
conditions on the currents corresponding to the dyon.  We found it
neccessary to impose  extra conditions, uniqueness of the groundstate and CP
invariance, in order to find  the dyon boundary state for $\theta =0$. The
states for general $\theta$ were then found by making the appropriate
unitary transformation which maps between the $\theta$-vacua in the bulk
gauge theory.  We note that the philosophy used here to determine the needed
boundary states is quite different than in the work on the Kondo effect.  In
that case physical considerations led to the ``fusion rule hypothesis''
which led to a unique boundary state.  In the dyon problem fusion has not
played a role.  Instead the current boundary conditions, together with a
groundstate uniqueness condition led to a one-parameter family of basis
states of dyon type. These involve Ishibashi states of a different type than
those that occur in the free or Kondo case due to the reversed sign in the
current boundary conditions.
 \section{Green's Functions} Now that we have determined the dyon boundary
state, we can calculate arbitrary Green's functions at long distances and
times compared to $r_0$ and $I$ using the boundary conformal field theory
techniques developed earlier.\cite{Cardy}  We shall see that this method
gives identical results to those obtained earlier by Polchinski\cite{Pol} by
integrating out the fermions using the Schwinger determinant formula.  This
provides a useful check on the validity of the boundary conformal field
theory technique for this problem.  We recall the basic ingredients of the
boundary conformal field theory calculation of Green's
functions.\cite{Cardy,CL,LA1,LA2}  The first, and crucial point is that left
and right Virasoro algebras are identified.  Furthermore, any Green's
functions involving only right-moving fields is unaffected by the boundary.
In the formulation of Section I this means that any Green's functions where
all points, $x_i$ are positive (or all negative) are unchanged.  In general,
Green's functions involving both left and right moving fields are treated as
if all fields were right-moving {}from the point of view of applying the
Virasoro and Kac-Moody algebras to simplify the form of the Green's
functions.  However, these algebras alone never completely determined the
Green's functions; certain additional conditions are needed.  Let us first
consider two-point functions. These must have the form:\cite{Cardy}
\begin{equation} <\Phi_R(z)\Phi_L^\dagger (z')> \propto
1/(z-z'^*)^{2x}\end{equation} The only undetermined feature is the overall
normalization.  This is determined by the boundary state.  In the case where
$\Phi_R$ is a Virasoro primary field with unit normalized two-point function
in the bulk: \begin{equation}  <\Phi_R(z)\Phi_R^\dagger (z')> =
1/(z-z')^{2x}, \end{equation} the boundary normalization is:\cite{CL}
\begin{equation} <\Phi_R(z)\Phi_L^\dagger (z')> ={<\Phi ;0|A> \over <0;0|A>}
1/[i(z^{'*}-z)]^{2x}\label{boundnorm}\end{equation}  Here $|A>$ is the
boundary state and $|\Phi ;0>$ is the product of left and right primary
states corresponding to the conformal tower of $\Phi$, which occurs in the
Ishibashi state of Eq. (\ref{Ishsol}).  $|0;0>$ is the vacuum  state.  In
the case of higher n-point Green's functions the algebras in general only
determine the Green's functions up to a larger number of parameters related
to the monodromy matrix.  These parameters can all be fixed by using the
operator product expansion together with Eq. (\ref{boundnorm}).

Let us first consider the fermion two-point functions.  We can immediately
restrict the possible non-zero two-point functions by using the $U(1)$ gauge
symmetry corresponding to $Q$ and the $SU(N)$ flavour symmetry.  In the
formulation of Section I, the charge of $\psi$ changes sign as it passes
through the origin.  Thus $<\psi^{i \dagger} (z)\psi_j(z^*)>$ vanishes due
to charge conservation.  (Throughout this section we use the convention that
$z=\tau+ir$ with $r\geq 0$.)   $<\psi_i (z_1)\psi_j(z_2^*)>$ is allowed to
be non-zero by charge conservation but vanishes by flavour symmetry for all
$N > 2$.  In the $N=2$ case it may have the form: \begin{equation} <\psi_i
(z_1)\psi_j(z_2^*)>={C \epsilon_{ij}\over z_1-z_2^*}\end{equation}  To
determine the normalization constant, $C$, we  use the boundary state.  In
this case the operator $\Phi_R$ in Eq. (\ref{boundnorm}) is $\psi_i(z_1)$,
the $p=Q=1$ primary field.  $\Phi_L^\dagger$ corresponds to
$\psi_j(-z_2^*)$.   The corresponding state does indeed occur in the dyon
boundary state $|D\theta >$ of Eq. (\ref{Deven}) and (\ref{abeven}).  The
needed amplitude is: \begin{equation} {<\Phi ;0 |A> \over<0;0|A>}={b_0\over
a_0}=e^{i\theta} \end{equation}  Using the standard bulk normalization for
the fermion two-point function we obtain  \begin{equation}
<\psi_i(z_1)\psi_j(z_2^*)> = {ie^{i\theta }\epsilon_{ij}\over 2\pi
(z_1-z_2^*)}\label{2ptfnc}\end{equation}  Imposing time-reversal symmetry
requires $\theta = 0$ or $\pi$.  Note that time reversal interchanges left
and right-movers.  In the present formulation it takes $\psi_i (t,r) \to
\psi_i (-t,-r)$.    Taking into account that time-reversal is anti-unitary
and setting the time co-ordinates to zero, we conclude: \begin{equation}
<\psi_1(r)\psi_2(-r)>=<\psi_1(-r)\psi_2(r)>^*\end{equation} This is only true
for $\theta = 0$ or $\pi$ {}from Eq. (\ref{2ptfnc}).  It was shown earlier that
the phase $\theta$ appearing in this Green's function should be identified
with the topological angle in the (3+1)-dimensional gauge theory. [See the
discussion in Ref. (\onlinecite{Pol})]. This topological term breaks
time-reversal symmetry except for these two values of $\theta$.  We note
that this Green's function is that of the free fermion theory with the
boundary condition of Eq. (\ref{N=2bc}).

Let us now turn to the four-point functions. These are determined by bulk
conformal field theory techniques together with Eq.
(\ref{boundnorm}).\cite{CL} [The corresponding calculations for the Kondo
problem appear in Ref. (\onlinecite{LA1}) and (\onlinecite{LA2}).]  We use
non-abelian bosonization to represent the fermion fields in terms of an
$SU(N)$ level $k=1$ Kac-Moody field $g_i$ and a free charge boson $\phi$.
We work entirely with right-movers on the whole real line.  $g_i$ can be
thought of as the right-moving part of an $SU(N)$ matrix field, $g_i^j$.
$\phi$ corresponds to the right-moving part of a free boson field. The
basic non-abelian bosonization formula is: \begin{eqnarray} \psi_i(r)
&\propto &g_i(r)e^{i\sqrt{4\pi /N}\phi (r)}  \nonumber \\ \psi_i(-r)
&\propto &g_i(-r)e^{-i\sqrt{4\pi /N}\phi (-r)} \ \ \label{NAB}\end{eqnarray}
(Here $r>0$.) The change in sign in the exponent for the charge boson
corresponds to the reversal of the charge of the fermion field upon passing
through the origin. With a simple free fermion boundary condition,
$\psi_R(0)=\psi_L(0)$, the bosonization formulas are the same but without
this sign change.

Green's functions in which all fields are on the same side of the origin are
unaffected by the impurity.  This is a general property of boundary conformal
field theory.\cite{Cardy}  Taking into account charge and flavour
conservation there are only two non-trivial four-point Green's functions,
$<\psi_i(z_1)\psi^{j \dagger}(z_2)\psi^{k \dagger}(z_3^*)\psi_l(z_4^*)>$ and
$<\psi_{i }(z_1)\psi_j(z_2)\psi_{k }(z_3^*)\psi_l(z_4^*)>$.  (The latter is
only non-zero for $N=4$.)  Using the non-abelian bosonization formula of
Eq.(\ref{NAB}), the first Green's function becomes: \begin{eqnarray}
<\psi_i(z_1)\psi^{j \dagger}(z_2)\psi^{k \dagger}(z_3^*)\psi_l(z_4^*)>
&=&<\exp\{i\sqrt{4\pi /N}[\phi (z_1)-\phi (z_2) \mp \phi (z_3^*)\pm \phi
(z_4^*)]\}>\nonumber \\ &&\cdot
<g_i(z_1)g^{j\dagger}(z_2)g^{k\dagger}(z_3^*)g_l(z_4^*)>\end{eqnarray} Here
the upper or lower signs refer to the case of free or dyon boundary
conditions respectively.  The free boson charge Green's function is given by:
\begin{eqnarray} &&<\exp\{i\sqrt{4\pi /N}[\phi (z_1)-\phi (z_2) \mp \phi
(z_3^*)\pm \phi (z_4^*)]\}>\propto  [(z_1-z_2)(z_3^*-z_4^*)]^{-1/N}\nonumber
\\ &&\ \ \cdot [(z_1-z_4^*)(z_2-z_3^*)/(z_1-z_3^*)(z_4^*-z_2)] ^{\pm
1/N}\label{chcorr}\end{eqnarray} where the upper and lower signs refer to the
free and dyon case respectively.  The WZW flavour Green's function is
determined by the Knizhnik-Zamolodchikov equations\cite{KZ} up to two free
parameters, $U_p$, $p=0,1$: \begin{equation}
<g_i(z_1)g^{j\dagger}(z_2)g^{k\dagger}(z_3^*)g_l(z_4^*)>=[(z_1-z_4^*)(z_2-z_3^*
)]^{-1+1/N}\sum_{p=0}^1 U_p[\delta^j_i\delta^k_l{\cal F}_1^{(p)}(x) +
\delta^j_l\delta^k_i{\cal F}_2^{(p)}(x)]\end{equation} Here $x$ is the
``cross-ratio'' \begin{equation} x\equiv {(z_1-z_2)(z_3^*-z_4^*)\over
(z_1-z_4^*)(z_3^*-z_2)}\end{equation} The functions ${\cal F}_A^{(0)}$ are
given by:\cite{KZ} \begin{eqnarray} {\cal
F}_1^{(0)}(x)&=&[x(1-x)]^{1/N-1}(1-x)\nonumber \\ {\cal
F}_2^{(0)}(x)&=&[x(1-x)]^{1/N-1}x\end{eqnarray} The other two functions can
be expressed in terms of hypergeometric functions.\cite{KZ}  As $x\to 0$
they both behave as: \begin{equation} {\cal F}_A^{(1)} \to
x^{1/N-1+N/(N+1)}\end{equation} We may take the proportionality in Eq.
(\ref{chcorr}) to be an equality by redefining the constants $U_p$.  By
choosing $U_0=-1/(2\pi )^2$, $U_1=0$, we obtain the correct free fermion
result: \begin{equation} <\psi_i(z_1)\psi^{j \dagger}(z_2)\psi^{k
\dagger}(z_3^*)\psi_l(z_4^*)> ={\delta^j_i\delta^k_l\over (2\pi
)^2(z_1-z_2)(z_3^*-z_4^*)}+{\delta^j_l\delta^k_i\over (2\pi )^2(z_1-z_3^*)
(z_4^*-z_2)}\end{equation}

Now let us consider the dyon case.  All that remains to be determined is the
two coefficients $U_p$.  These can be fixed by considering the bulk limit,
$z_1 \to z_2$, $z_3^*\to z_4^*$.  Note that the second factor in Eq.
(\ref{chcorr}), whose exponent takes the opposite sign in the dyon case,
goes to $(-1)^{\pm 1/N}$ in this limit.  Therefore, in order to obtain the
free result in this bulk limit, we must choose the same values for the
parameters, $U_p$.  In particular we must again choose $U_1=0$ in order to
avoid a singularity of the form $[(z_1-z_2)(z_3^*-z_4^*)]^{N/(N+1)-1}$ since
this does not correspond to the dimension of any bulk operator.  Thus we
obtain: \begin{eqnarray} &&<\psi_i(z_1)\psi^{j \dagger}(z_2)\psi^{k
\dagger}(z_3^*)\psi_l(z_4^*)> =\left[{(z_1-z_3^*)(z_4^*-z_2)\over
(z_1-z_4^*)(z_2-z_3^*)}\right]^{2/N} \nonumber \\ && \ \ \cdot
\left[{\delta^j_i\delta^k_l\over (2\pi
)^2(z_1-z_2)(z_3^*-z_4^*)}+{\delta^j_l\delta^k_i\over (2\pi )^2(z_1-z_3^*)
(z_4^*-z_2)}\right]\label{moncorr}\end{eqnarray}

 By now considering the boundary limits of this expression we can deduce the
presence of various boundary operators in the operator product expansion.
These operators must correspond to the states in the spectrum with dyon
boundary conditions at both ends, $Z_{D\theta ,D\theta}$.\cite{Cardy}
Choosing equal angles $\theta$ at both ends the spectrum of primary states
corresponds to scaling dimensions: \begin{equation} x=Q^2/2N
+p(N-p)/2N\end{equation} with the restriction: \begin{equation} p=Q\
(\hbox{mod}\  2)\end{equation} Let us now consider the surface limit $z_1
\to z_3^*$, $z_2 \to z_4^*$ of Eq. (\ref{moncorr}).  (In this limit all four
points approach the boundary.)  This gives: \begin{equation}
<\psi_i(z_1)\psi^{j \dagger}(z_2)\psi^{k \dagger}(z_3^*)\psi_l(z_4^*)> \to
{\delta^j_l\delta^k_i \over (2\pi )^2
[(z_1-z_3^*)(z_4^*-z_2)]^{1-2/N}|\tau_1-\tau_2|^{4/N}}\end{equation} This
leading singular behaviour results {}from the boundary OPE: \begin{equation}
\psi_i(z_1)\psi^{k \dagger}(z_3^*)\to {\delta^k_i\over (2\pi
)(z_1-z_3^*)^{1-2/N}}{\cal O}_{2,0}(\tau_1)\end{equation} where ${\cal
O}_{2,0}(\tau_1)$ is the $Q=2$ $p=0$ charge 2 flavour singlet boundary
operator of dimension $x=Q^2/2N=2/N$ and correlation function:
\begin{equation} <{\cal O}_{2,0}(\tau_1){\cal
O}_{2,0}(\tau_2)>=|\tau_1-\tau_2|^{-4/N}\end{equation} The other boundary
limit is $z_1\to z_4^*$, $z_2\to z_3^*$.  This gives: \begin{equation}
<\psi_i(z_1)\psi^{j \dagger}(z_2)\psi^{k \dagger}(z_3^*)\psi_l(z_4^*)> \to
{(-\delta^j_i\delta^k_l+\delta^j_l\delta^k_i)\over (2\pi
)^2[(z_1-z_4^*)(z_2-z_3^*)]^{2/N} |\tau_1-\tau_2|^{2-4/N}} \end{equation}
In this case the boundary OPE is: \begin{equation}  \psi_i(z_1)\psi_l(z_4^*)
\to {\cal O}_{0,2,il}(\tau_1)/(2\pi )(z_1-z_4^*)^{2/N} \end{equation} where
${\cal O}_{0,2,il}(\tau_1)$ is the charge 0 $SU(N)$ antisymmetric tensor,
$Q=0$, $p=2$, field of dimension $x=p(N-p)/2N=1-2/N$ and correlation
function: \begin{equation} <{\cal O}_{0,2,il}(\tau_1){\cal
O}_{0,2}^{kj\dagger}(\tau_2)>
=(\delta^k_i\delta^j_l-\delta^j_i\delta^k_l)/|\tau_1-\tau_2|^{2-4/N}
\end{equation}

Now we turn to the other non-trivial four-point function,
$<\psi_i(z_1)\psi_j(z_2)\psi_k(z_3^*)\psi_l(z_4^*)>$.  This must vanish by
$SU(N)$ symmetry except for $N=4$ where it may be proportional to the
antisymmetric tensor invariant $\epsilon_{ijkl}$.  Note that this operator
has charge $Q=0$ since two points are on the positive axis and two on the
negative axis.  Unlike the previous case, this Green's function vanishes in
the free fermion case due to charge conservation.  In the dyon case, with
$N=4$, using non-abelian bosonization it becomes: \begin{eqnarray}
<\psi_i(z_1)\psi_j(z_2)\psi_k(z_3^*)\psi_l(z_4^*)>&=&<\exp \{i\sqrt{\pi }
[\phi (z_1)+\phi (z_2)-\phi (z_3^*)-\phi (z_4^*)]\}>\nonumber \\ && \ \
<g_i(z_1)g_j(z_2)g_k(z_3^*)g_l(z_4^*)>\end{eqnarray} The charge correlation
function is: \begin{equation} <\exp\{i\sqrt{\pi }[\phi (z_1)+\phi (z_2)-\phi
(z_3^*)-\phi (z_4^*)]\}> \propto \left[{(z_1-z_2)(z_3^*-z_4^*)\over
(z_1-z_3^*)(z_1-z_4^*)(z_2-z_3^*)(z_2-z_4^*)}\right]^{1/4}\end{equation} The
flavour correlation function must be a solution of the appropriate
Knizhnik-Zamolodchikov (KZ) equation.  In this case there is only a single
allowed tensor structure, $\epsilon_{ijkl}$ rather than two as in the
previous case.  Consequently the KZ equation should only have one solution.
This solution can be determined by dimensional analysis and the symmetry
between the four co-ordinates to be: \begin{equation}
<g_i(z_1)g_j(z_2)g_k(z_3^*)g_l(z_4^*)>\propto
\epsilon_{ijkl}/[(z_1-z_2)(z_3^*-z_4^*)
(z_1-z_3^*)(z_1-z_4^*)(z_2-z_3^*)(z_2-z_4^*)]^{1/4}\end{equation} Combining
these two factors we obtain: \begin{equation}
<\psi_i(z_1)\psi_j(z_2)\psi_k(z_3^*)\psi_l(z_4^*)>=C\epsilon_{ijkl}/
[(z_1-z_3^*)(z_1-z_4^*)(z_2-z_3^*)(z_2-z_4^*)]^{1/2}\end{equation} where $C$
is a constant, to be determined.  Note that we have determined the form of
the Green's function by symmetry and scaling considerations up to a single
constant, $C$, which can be determined {}from the boundary state.  This is
done by taking the bulk limit, $z_1\to z_2$, $z_3^*\to z_4^*$ in which:
\begin{equation} <\psi_i(z_1)\psi_j(z_1)\psi_k(z_3^*)\psi_l(z_3^*)>=
C\epsilon_{ijkl}/ (z_1-z_3^*)^2\end{equation} Note that this limit is given
by the bulk OPE: \begin{equation} \psi_i(z_1)\psi_j(z_2) \to (1/2\pi ){\cal
O}_{2,2,ij}(z_1)\end{equation} where ${\cal O}_{2,2,ij}$ is the charge 2
antisymmetric tensor $Q=p=2$ operator of dimension $x=1$ with bulk
correlation function: \begin{equation} <{\cal O}_{2,2,ij}(z_1){\cal
O}_{2,2}^{kl\dagger}(z_2)>_{\hbox{bulk}}=(\delta_i^k\delta_j^l-\delta_j^k\delta_i^l)
/(z_1-z_2)^2 \end{equation} To determine the constant, $C$, we need the
two-point function of ${\cal O}_{2,2,ij}$ when the two points straddle the
boundary.  This depends on the boundary state, $|D\theta >$.  The state
corresponding to the operator ${\cal O}_{2,2,ij}(z){\cal O}_{2,2,kl}(z^*)$ is
contained in the Ishibashi state $|2>_{c-}\otimes |2>_f$.  Thus the constant
C is determined {}from  $b_0/a_0=e^{i\theta}$ {}from Eq. (\ref{Deven}) and
(\ref{abeven}).  \begin{equation}
<\psi_i(z_1)\psi_j(z_2)\psi_k(z_3^*)\psi_l(z_4^*)>={e^{i\theta}\epsilon_{ijkl}
\over (2\pi
)^2[(z_1-z_3^*)(z_1-z_4^*)(z_2-z_3^*)(z_2-z_4^*)]^{1/2}}\label{moncorr2}\end{equation}
Again, time-reversal invariance requires $\theta=0$ or $\pi$; $\theta$ is
identified as the  topological angle of the bulk gauge theory.

The boundary limit, $z_1\to z_3^*$, is determined by the boundary OPE:
\begin{equation} \psi_i(z_1)\psi_k(z_3^*) \to {e^{i\theta /2}{\cal
O}_{02,ik}(\tau_1)\over 2\pi (z_1-z_3^*)^{1/2}} \end{equation} where ${\cal
O}_{02,ik}$ is the $Q=0$, $p=2$ operator of $x=1/2$, and correlation
function: \begin{equation} <{\cal O}_{02,ik}(\tau_1){\cal
O}_{02,jl}(\tau_2)>= {\epsilon_{ijkl}\over |\tau_1-\tau_2|}\end{equation}

Both results, Eq. (\ref{moncorr}) and (\ref{moncorr2}) agree exactly with
those obtained by Polchinski\cite{Pol} by integrating out the fermions.  It
seems clear that higher n-point correlation functions can also be obtained
using the present methods.  In particular, consider non-zero Green's
functions with a total charge of $nN/2$ at $x>0$ (and hence  charge $-nN/2$
at $x<0$). The corresponding boundary matrix element is proportional to
$a_{n/2}$ for $n$ even or $b_{(n-1)/2}$ for $n$ odd. We see {}from Eq.
(\ref{abeven}) that this Green's function has the $\theta$-dependence
$e^{in\theta}$ in agreement with Polchinski.\cite{Pol} \section{Conclusions}
We conclude by comparing the Kondo and Callan-Rubakov effects, {}from the
viewpoint of boundary conformal field theory.  Both problems involve free
fermions interacting with a localized quantum mechanical degree of freedom.
Both problems involve singular s-wave scattering effects; since higher
spherical harmonics are unimportant, they can be reduced to one-dimensional
problems involving incoming and outgoing waves on the half-line or
equivalently right-movers only on the entire line. In both problems it is
convenient to use non-abelian bosonization to separate charge and spin (or
more generally flavour) degrees of freedom.  In the Kondo problem, the
interaction only involves spin (ie. flavour) fields; in the dyon problem it
only involves charge fields.  In both problems the local quantum-mechanical
degree of freedom can be eliminated {}from the low-energy, long-wavelength
effective theory, leaving behind only a boundary condition.  In the Kondo
problem this elimination can be affected by redefining the spin current so
as to ``adsorb'' the impurity spin.  In the dyon problem, a new
gauge-invariant charge current is defined which involves the dyon as well as
the fermions.  (This is the analogue of the adsorption process.)   In the
Kondo problem the effective boundary condition can be found {}from the fusion
rules.  This doesn't seem to be possible for the dyon problem where a
different class of Ishibashi states, dictated by the reversed sign of the
current boundary conditions, occurs.

Renormalization group ideas play a crucial role in our understanding of the
Kondo effect.  The effective theory with the impurity eliminated and an
effective boundary condition should be regarded as an infrared stable fixed
point to which the system renormalizes.  The dyon problem is effectively
harmonic in bosonic variables and there is not really any coupling constant
to renormalize. As far as we can see, renormalization group ideas don't seem
to be relevant.  Nonetheless, there is an effective description, valid at
sufficiently long distances and low energies, with the impurity eliminated
and replaced by an effective conformally invariant boundary condition.

There are versions of both problems for which the effective boundary
conditions can be expressed linearly in the fermion fields.  These may be
thought of as ``Fermi liquid boundary conditions''.  (In the dyon problem,
this corresponds to $N\leq 2$.) In other cases the boundary conditions are
of ``non-Fermi liquid'' type and lead to exotic behaviour such as fractional
scaling dimensions and fractional particle production .

\section{Acknowlegements} We would like to thank C.G. Callan, A.W.W. Ludwig,
J. Polchinski and E. Wong for helpful discussions.  IA thanks the Institut
des Hautes Etudes Scientifiques, Bures-sur-Yvette, France for hospitality
while some of this work was carried out.  This research was supported in
part by NSERC of Canada.

 \end{document}